\documentstyle[aps,epsfig]{revtex}
\draft
\def\bm#1{\mbox{\boldmath$#1$}}
\newcommand{\jone}{\mbox{${\cal J}^{(1)}$}}

\begin{document}
\title{Quadrupole pairing interaction and signature inversion}
\author{F.R. Xu$^{1,2}$, W. Satu{\l}a,$^{1,3}$ and R. Wyss$^{1,4}$}
\address{$^1$Royal Institute of Technology,
             Physics Department Frescati,\\
             Frescativ{\"a}gen 24, S-104 05 Stockholm, Sweden}
\address{$^2$Department of Technical Physics,
             Peking University, Beijing 100871, China}
\address{$^3$Institute of Theoretical Physics, University of Warsaw,
             ul. Ho{\.z}a 69, PL-00 681 Warsaw, Poland}
\address{$^4$Department of Technology, Kalmar University, Box 905, 391 29 Kalmar}

\maketitle

\begin{abstract} 
The signature inversion in the
$\pi h_{11/2}\bigotimes\nu h_{11/2}$ rotational bands of
the odd-odd Cs and La isotopes and the $\pi h_{11/2}\bigotimes\nu i_{13/2}$
bands of the odd-odd Tb, Ho and Tm nuclei is investigated using
pairing and deformation self-consistent mean-field calculations.
The model can rather satisfactorily 
account for the anomalous signature 
splitting provided that spin assignments in some of the bands 
are revised. Our  calculations show that signature inversion 
can appear already at axially symmetric shapes.
It is found that this is due to the contribution of 
the $(\lambda\mu)=(22)$ component of the quadrupole-pairing interaction
to the mean-field potential.

\end{abstract}


\section{Introduction}

The invariance of the intrinsic Hamiltonian with respect to
the signature symmetry
gives rise to the occurrence of two rotational bands in
odd and odd-odd nuclei that differ in spin
by $1\hbar$. These bands are built upon
the same intrinsic configuration but differ in their signature symmetry
eigenvalue, $r$~\cite{[Boh78]}. Often the signature exponent $\alpha$ 
[$I \equiv \alpha $~mod~2] is used to label these bands where $\alpha$ is
related to $r$ via  $r=e^{-i\pi\alpha}$.

In general, signature partner bands are not equivalent energetically.
The energy difference
is called signature splitting when measured in the
rotating frame of reference as a function of frequency.
The origin of the splitting is essentially due to the mixing of 
the $\Omega=1/2$ state into wave function. Since this component has a 
non-zero diagonal matrix element of the cranking operator $\omega\hat j_x$ of 
opposite sign for each signature, one expects one signature  
to gain (or lose) energy with increasing frequency with respect to
the other. 
For high-$j$ one-quasiparticle (1-qp) unique-parity configurations, 
the favored signature is obtained by the simple rule, 
$\alpha_f = (-)^{j-1/2}1/2$~\cite{[Ste75]}. 
However, in unique parity 2-qp configuration in odd-odd nuclei, it may occur
that $\alpha_{f} = (-)^{j_\nu-1/2}1/2+ (-)^{j_\pi-1/2}1/2$ band may become
energetically unfavored.
Similar effects have been observed in some odd-A nuclei in 3-qp 
configuration~\cite{[Ham86]}.
This phenomenon is called signature inversion~\cite{[Ben84]}.

Signature inversion has been observed, for example, in odd-odd nuclei for
the $\pi h_{11/2}\bigotimes\nu h_{11/2}$ configurations in 
the A$\sim$130 and $\pi h_{11/2}\bigotimes\nu i_{13/2}$
configurations in A$\sim$160 regions.
Different theoretical attempts have been presented to
interpret the phenomenon. Triaxiality of the nuclear shape was suggested in 
Refs.~\cite{[Fra83],[Ham83],[Ham83b],[Ben84]}.  
The general condition for signature inversion to take place~\cite{[Ben84]} 
was positive triaxial deformation, $\gamma >0$, within the  Lund convention 
combined with
a specific position of the Fermi surface, see also~\cite{[Ike88],[Mat89]}. 
One should point out, however, that {\bf (i)} the $\gamma$-values
are almost impossible to measure directly in experiment; 
{\bf (ii)} the values of the $\gamma$ parameter necessary to account for the
experimental data are not always supported by potential energy calculations 
which in many cases predicts almost axial shapes \cite{[Ced92]}.
Triaxiality seems, therefore, to be not
always sufficient calling for alternative or supplementary mechanism. 

Hamamoto~\cite{[Ham86b]} and Matsuzaki~\cite{[Mat91]} analyzed 
the long-range, residual proton-neutron ($pn$) $\chi Q_p Q_n$  interaction.
They concluded that the
self-consistent value of the strength parameter $\chi$ is far too weak to 
account for the empirical data.
Semmes and Ragnarsson~\cite{[Sem91]} considered a residual
$pn$ contact force with spin-spin interaction,
$V_{pn}\propto \delta( \bm{r}_p - \bm{r}_n)
(u_0 + u_1 \vec{\sigma}_n \cdot  \vec{\sigma}_p)$, 
in the framework of the particle-rotor model.
Remarkably, using one set of parameters 
$u_0$, $u_1$, it appeared possible to reproduce very accurately some
 $\pi h_{11/2}\bigotimes\nu h_{11/2}$ bands in the A$\sim$130 region
 and $\pi h_{11/2}\bigotimes\nu i_{13/2}$ bands in rare-earth nuclei,
without invoking triaxiality~\cite{[Ced92],[Taj94]}.
However, to account for the 
$\pi h_{9/2}\bigotimes\nu i_{13/2}$ bands in rare-earth nuclei,
a substantial reduction of the strength parameters 
$u_0$ and $u_1$ was necessary~\cite{[Bar97]}.  
Signature inversion is also obtained
in the projected shell-model at axial symmetric shapes, 
resulting from the crossing of
different bands with opposite signature dependence\cite{[Har91]}.

Another probe of the mechanism behind signature inversion are
the electromagnetic transition amplitudes. The ratios of the
BM1/BE2 transition rates 
are very sensitive to the deformation parameters ($\gamma$ and $\beta$) 
and  whether the decay goes from favored states to unfavored 
or vice versa\cite{[Ham86]}. 
In the present paper, however, we restrict ourselves entirely
to the systematics of Routhians and spins. 
Calculations of BM1/BE2 ratios are beyond
the scope of this work.

In Ref.~\cite{[Sat96]} we have published the results of 
deformation and pairing self-consistent Total Routhian Surface (TRS) 
calculations for $^{120}$Cs.  
It has been demonstrated that the inclusion of quadrupole pairing 
correlations ($QQ$-pairing), number projection and a proper treatment 
of blocking
into the TRS model made it possible to reproduce signature inversion
without invoking the $pn$ interaction. In this paper, we demonstrate
that our extended TRS model is able to reproduce rather accurately
and systematically the signature inversion in odd-odd 
$^{120-124}$Cs, $^{124-128}$La, $^{154-158}$Tb, $^{156-160}$Ho and 
$^{158-162}$Tm nuclei and that
the mechanism causing signature inversion  
is triaxiality combined with the contribution of the $(\lambda\mu ) =(22)$ 
component of the $QQ$-pairing force to the single-particle potential.

In the next section, we present a brief outline of our model.
A general discussion on the role of the $QQ$-pairing
force with respect to signature
inversion is given in Sec.~III. 
The discussion of 
empirical spin assignments and the results of self-consistent TRS
calculations 
are presented in Sec.~IV
for the odd-odd 
$^{120-124}$Cs, $^{124-128}$La, $^{154-158}$Tb, $^{156-160}$Ho and 
$^{158-162}$Tm nuclei. 
A summary is given in the last section.

\section{The model}

The TRS method\cite{[Naz89]} is a macroscopic-microscopic approach 
in a uniformly-rotating body-fixed frame of reference~\cite{[Ing54]}. 
The total Routhian of a nucleus is calculated 
on a grid in deformation space,  using the Strutinsky shell correction 
method~\cite{[Str66]}. The model employs the deformed 
Woods-Saxon potential of Ref.~\cite{[Cwi87]} and liquid-drop model
of Ref.~\cite{[Coh74]}.
The pairing energy is calculated using 
a separable interaction of seniority 
and doubly-stretched quadrupole type:
\begin{equation}
\bar{v}_{\alpha\beta\gamma\delta}^{(\lambda\mu)} = - G_{\lambda\mu}
g_{\alpha\bar\beta}^{(\lambda\mu)}
g^{*(\lambda\mu)}_{\gamma\bar\delta}\;,
\end{equation}
where
\begin{equation}
g_{\alpha\bar\beta}^{(\lambda\mu)}=\left\{
\begin{array}{ll}
\delta_{\alpha\bar\beta}\;, & \lambda=0, \quad \mu=0 \\
\langle \alpha | \hat Q_{\mu}^{\prime\prime} |\bar\beta\rangle, &
\lambda=2, \quad \mu=0,1,2.
\end{array}
\right.
\end{equation}
The above expression employs good signature basis~\cite{[Goo74]},
$\bar\alpha =\hat T\alpha$ 
stands for the time-reversed state and $\overline{|r=\pm i\rangle} = 
\pm |r=\mp i\rangle$.
To avoid the sudden collapse
of pairing correlations, we use the Lipkin-Nogami 
approximate number projection~\cite{[Pra73]}.
It is important to stress that the model is {\it free\/} of adjustable 
strength parameters. The monopole pairing strength, $G_{00}$, is 
determined by the average gap method of Ref.~\cite{[Mol92]} and the 
$QQ$-pairing strengths, $G_{2\mu}$, are calculated 
to restore the Galilean invariance broken by the seniority pairing
force, see
the prescription in Ref.~\cite{[Sak90]}. 
For example, in the A$\sim$130 mass region, where the 56(66) lowest
proton (neutron) single-particle levels are used for the pairing
calculation, the pairing strengths parameters are equal
$G_{00}\approx 165(145)$\,keV and
$G_{2\mu}\approx 4.4(3.3)$\,keV/fm$^4$ for protons (neutrons).
In the A$\sim$160 mass region,  single-particle levels between
10-70(15-90) were used for the pairing calculations for protons (neutrons)
and typical values of the strength parameters are
$G_{00}\approx 140(105)$\,keV and $G_{2\mu}\approx 3.0(1.6)$\,keV/fm$^4$
for protons (neutrons). Let us recall that the
use of the double-stretched generators for the $QQ$-pairing force
result in, to a large extent,
isotropic and shape independent values of
$G_{2\mu}$~\cite{[Sat94b]}.

The resulting cranked-Lipkin-Nogami (CLN) equation takes the form of the
well known Hartree-Fock-Bogolyubov-like (HFB) equation. In the TRS model,
the CLN equation is solved self-consistently at each frequency and 
each grid point in deformation space which includes quadrupole
$\beta_2, \gamma$ and hexadecapole $\beta_4$ shapes
(pairing self-consistency). Finally, the equilibrium deformations 
are calculated by minimizing  the total Routhian with 
respect to the shape parameters (shape self-consistency).
For further details concerning the formalism, we refer the reader to
Ref.~\cite{[Sat94],[Sat94a],[Sat95]}.

\section{The influence of the 
quadrupole pairing force on signature inversion}

In our previous works~\cite{[Sat94a],[Wys95]}, it has been shown
that $QQ$-pairing, in particular its time-odd $\Delta_{21}$ component, 
plays an important role in the alignment of quasi-particles. It strongly
influences the moment of inertia and partly can account for twinning  
of the spectra of odd and even super-deformed  nuclei in the A$\sim$190 mass 
region.
The two-body pairing interaction enters also the single-particle 
channel via the $\Gamma$ potential:
\begin{equation}\label{gam}
\Gamma_{\alpha \beta}^{(\lambda\mu)} = -
G_{\lambda\mu}\displaystyle\sum_{\gamma\delta>0}
g_{\alpha\bar\gamma}^{(\lambda\mu)} g_{\beta\bar\delta}^{*(\lambda\mu)}
\rho_{\delta\gamma}\;,
\end{equation}
where $\rho$ is the density matrix. The single-particle
contribution coming
from the separable pairing force 
is usually small and in many applications
simply disregarded. However, in the Lipkin-Nogami approach, 
it has to be taken into account for the consistency of the method.
Interestingly, in spite of its weakness, the  $\Gamma^{(22)}$ field
plays a rather important role for the signature inversion as will be 
demonstrated in the following.

Let us consider an odd-odd nucleus with an unpaired proton and 
neutron occupying 
the high-$j$, unique-parity orbits. The favored signature is expected to be
$\alpha_f = (-)^{j_\pi -1/2}1/2 + (-)^{j_\nu -1/2}1/2$. However, when one of 
the particles is low in the shell 
and the other occupies orbits around the middle or above the middle of 
the shell,
signature inversion can occur. 
In the A$\sim$130 region e.g.,
two signatures, $\alpha_{f(uf)} = (\alpha_\pi=-1/2)\otimes(\alpha_\nu=\mp 1/2) 
= 1(0)$, of the  
 $\pi h_{11/2}\bigotimes\nu h_{11/2}$ configuration
are seen in experiment.
For the protons, only the signature-favored $\alpha_\pi=-1/2$
orbit is observed since 
the signature splitting of  $\pi h_{11/2}$ low-$\Omega$
levels is typically several hundred keV.
For instance, in $^{120}$Cs, it is calculated to be more than 0.7\,MeV
at $\hbar\omega=0.2$\,MeV.
The role of this low-$\Omega$, high-$j$ orbit is 
to induce a deformation close to axial symmetry with slight tendency
towards positive $\gamma$ value~\cite{[Ben84]}.
Otherwise it acts as spectator and the signature splitting is
entirely due to the neutrons.

Fig.~1 shows the influence of each $Q_{2\mu}$
($\mu=0, 1, 2$) component of the $QQ$-pairing interaction on
the signature splitting 
$\Delta e^\prime \equiv e^{(uf,\alpha =0)} - e^{(f,\alpha =1)} $, 
between the unfavored $e^{(uf,\alpha =0)}$ and favored
$e^{(f,\alpha =1)}$ Routhians as a function of the 
quadrupole pairing strength
$G_{2\mu}$ relative to the value $G^{sc}_{2\mu}$ determined
according to Ref.~\cite{[Sak90]}.  
The calculations were performed for
$^{120}$Cs at fixed frequency and axial shape in order to address
the effects of the $QQ$-pairing force alone.  
To find the origin of the signature inversion,
we performed calculations with and without the 
mean-field contributions $\Gamma^{(2\mu)}$.
The figure nicely demonstrates the role of the $\Gamma^{(22)}$ potential:
It is the only component of ($\lambda\mu$) that creates 
signature inversion of the order of a few tens of keV
already at axial shape.

To gain a better understanding of the role played by
the mean-field potential $\Gamma^{(2\mu)}$ as a function
of neutron number, Fig.~2 shows the
contribution to the signature splitting 
steming from the $\Gamma$-potential
$\Delta e_\Gamma = e^{(uf,\alpha = 0)}_{\Gamma}
- e^{(f, \alpha = 1)}_{\Gamma}$, where:
\begin{equation}
e^{(\alpha)}_{\Gamma} = {1\over 2} \bm{Tr} ( \Gamma^{(\alpha)}\rho^{(\alpha)} )
\end{equation}
as a function of $N$ in the A$\sim$130 mass region
(superscript $(\alpha)$ refers to the signature of the blocked 
$h_{11/2}$ quasi-particles).   
Neutron number N=59 corresponds to the occupation of
the $\Omega=1/2$ orbit, 
while at N=75 the $\Omega=9/2$ orbit of the $h_{11/2}$ subshell becomes 
occupied. Note again, that axial, fixed deformation was chosen to address  
the role of the $QQ$-pairing force effects alone.

Obviously, the contribution $\Delta e_\Gamma$ varies with 
the position of the Fermi energy.
There is a clear tendency for all components of the pairing force to
favor the unfavored signature.
At the bottom of the $\nu h_{11/2}$ subshell
(corresponding to 59$\leq$N$\leq$61)
the $\Gamma^{(20)}$, $\Gamma^{(21)}$ and even $\Gamma^{(00)}$
show trends
(negative contributions) towards the anomalous signature splitting.
In these cases, however, the 
signature splitting caused by the Coriolis force is
considerably larger, implying that signature inversion cannot
occur. Around the middle of the shell (corresponding to 63$\leq$N$\leq$69) 
where  the signature splitting induced by the Coriolis force
is rather small or even absent at low rotational frequencies,
the $\Gamma^{(22)}$ potential favors the inversion
and may even compete with the Coriolis force.
Note, that this corresponds to the neutron numbers where indeed 
the signature inversion has been observed
experimentally, see e.g. Ref.~\cite{[Kom93]} and references therein.

The signature splitting,  $\Delta e^\prime$, as a function of the neutron 
shell filling with and without $QQ$-pairing is shown 
in Fig.~3. 
The calculations were done 
at fixed {\it triaxial\/} shape of $\gamma=15^\circ$ and at 
$\hbar\omega = 0.2$\,MeV. Both sets of the calculations result in signature 
inversion for neutron numbers $\ge 63$.
However, a clear increase in the anomalous splitting
of the order $\sim$40\,keV is caused by the
$QQ$-pairing  force. 
The size of the signature inversion as a function of the triaxiality
parameter $\gamma$ for the case of 
$^{120}$Cs can be studied in Fig.~4. 
Note, that the contribution due to the $QQ$-pairing 
is almost independent of $\gamma$ due to
double stretching~\cite{[Sat93]}. 
In a previous work~\cite{[Ced92]} on $^{120}$Cs, the
$\gamma$ deformation obtained from the earlier TRS calculations 
(including only seniority
pairing treated non-self-consistently and no number projection)
could not account for the observed signature splitting. 
In the extended TRS calculations, the combined effects of
triaxiality and $QQ$-pairing force reproduce the 
data in this nucleus very well~\cite{[Sat96]}.

For the A$\sim$130 and A$\sim$160 nuclei, the anomalous signature splitting
happens at low rotational frequencies where the Coriolis mixing is weak. 
However, with increasing rotational frequency, the Coriolis force dominates
always and restores signature splitting to normal order.
The critical frequency, $\hbar\omega_c$, at which this takes place
is another characteristic quantity of signature inversion.
Fig.~5 shows a significant increase in $\hbar\omega_c$
due to the presence of $QQ$-pairing.
Finally, Fig.~6 shows the effect of the $QQ$-pairing
force on the total angular momentum
$I_x$ as a function of the shell filling. The calculations were performed
at $\hbar\omega=0.2$\,MeV. In general,
$QQ$-pairing slightly increases (decreases) $I_x$ for the unfavored (favored)
signature, respectively. Moreover, the effect is somewhat
larger for the unfavored signature branch.
Although globally the effect is rather modest, it clearly reduces
the normal signature splitting or enhances anomalous signature splitting,
depending on neutron number.

Let us make the following few remarks summarizing the discussion of this 
section. It seems rather well documented that signature inversion 
(in axially deformed nuclei) in
our calculation is due to the $\Gamma^{(22)}$ potential. The effect is
not accidently related to the particular choice of pairing strength parameters.
Indeed, a change of $G_{00}$ and(or) $G_{2\mu}$ (see Fig.~1) by$\pm$10\%
only weakly affects the calculated value 
of $\Delta e^\prime$. Our calculations also do not show that
the effect is related to number-projection. Analogous
calculations but performed within the BCS approximation 
led us to similar conclusions. However, it is of crucial importance to 
perform rigorous blocking for each signature separately. Indeed, 
approximating the odd-N system by a one quasiparticle state
created on top of the odd-N vacuum  does not result in any inversion. 
Blocking of signature partners at non-zero rotational frequency
(where signature splitting due to time-reversal symmetry breaking sets in)
leads to, in general small, differences in density matrices
$\rho ^{(\alpha_f)}$ and $\rho^{(\alpha_{uf})}$ which in 
turn result in differences between
$\Gamma^{(\lambda\mu)}_{\alpha_f}$ and 
$\Gamma^{(\lambda\mu)}_{\alpha_{uf}}$
potentials for favored and unfavored signature bands, respectively.
However, we were not able to recognize a simple mechanism causing 
the small differences in the density matrices to add up "coherently" 
and eventually form  such a regular pattern as depicted in Figs.~1-6.

\section{The TRS results: Comparison with experiment for the A$\sim$130 
and A$\sim$160 nuclei}

In this section, we present the results from
the pairing and deformation self-consistent TRS model
described in Sec.~II
for the odd-odd 
$^{120-124}$Cs, $^{124-128}$La, $^{154-158}$Tb, $^{156-160}$Ho and 
$^{158-162}$Tm nuclei. In order to compare theory and experiment,
the correct spin assignments of the bands are crucial. 
Indeed, by changing the spin values
by one unit, the signature splitting becomes inverted and
a totally different pattern emerges. Particularly in odd-odd nuclei, 
due to the complexity of the low-spin spectra, 
spins are not always directly determined in 
experiment. This is the case for many
of the rotational bands
associated with the $\pi h_{11/2}\bigotimes\nu h_{11/2}$
configuration in the $A\sim 130$ region and
$\pi h_{11/2}\bigotimes\nu i_{13/2}$ configurations in the $A\sim 160$ region.
Therefore, we start our discussion by revisiting current   
spin assignments in these bands.

As mentioned above, the spin assignments of many rotational bands
associated with the $\pi h_{11/2}\bigotimes\nu h_{11/2}$
configuration in the $A\sim 130$ region and
with the $\pi h_{11/2}\bigotimes\nu i_{13/2}$ in the $A\sim 160$ region
are tentative and based mainly on systematics. However, if the underlying 
assumptions or "the first guesses" 
are false, incorrect assignment may spread over many nuclei.
Indeed, the spin assignments in these nuclei are very controversial.
In $^{128}$La, for example, the recent experiment~\cite{[Hay95]}
firmly established the experimental spins via in-beam and
$\beta$-decay measurements.  The previous
spin assignment of Ref.~\cite{[God89]} was lower by $3\hbar$!

Our calculation is in good agreement with the experimental data
in $^{128}$La based on the recent spin assignment~\cite{[Hay95]}, 
see Fig.~8. Also for 
$^{120}$Cs~\cite{[Ced92]}, very good agreement between calculations and
experiment is achieved,
see Fig.~8 and~\cite{[Sat96]}.
Therefore, we choose these two nuclei as the reference nuclei
to verify spin assignments in the $\pi h_{11/2}\bigotimes\nu h_{11/2}$
bands in the Cs and La isotopes. We further assume in our analysis
that nuclear moments of inertia \jone$\equiv I_{x}/\omega(I)$
smoothly decrease with increasing neutron number.
This assumption is supported by comparison with neighboring 
even-even nuclei and by deformation systematics\cite{[Wys89]},
see also Fig.~10.
Based on the systematic analysis and at the same time compared with 
our cranking calculations, new spin values are suggested as shown in Table~I.

The experimental \jone~moments of inertia for the Cs and La isotopes 
corresponding to our 
new assignments are shown in Fig.~7. They decrease smoothly with 
increasing Fermi energy. The sensitivity of the method is shown for
the case of $^{130,132}$La. As is seen in Fig.~7,
changing the spins by $\pm 1\hbar$ introduces rather sharp
kinks in the sequence of the moments of
inertia in these nuclei, imposing rather strong restrictions for the 
relative spin values between the different isotopes.
This analysis can of course not replace any experimental verification,
but at present corresponds to the most reliable way
in the absence of accurate data.

We have also investigated the spin assignments for the
$\pi h_{11/2}\bigotimes\nu i_{13/2}$ bands of $^{154,156}$Tb,
$^{156-160}$Ho and $^{158-162}$Tm.
For this mass region,
our calculations agree well with the bulk part of the data
as well as the recent analysis\cite{[Liu95]}.
Systematics of the empirical moments
of inertia (\jone~is expected to increase with neutron number
in this mass region)
and spins suggest that the experimental assignments 
of $^{158}$Ho and $^{156}$Tb are too low.
Such assignments can in general not account for the initial alignment 
that one expects from low-$\Omega$, high-$j$ orbits.
With the guidance of
the available spin assignments and our cranking calculations,
we suggest the spin values presented in Table I.
The comparison between experimental (with our spin assignments) and 
theoretical values of $I_x$ versus $\hbar\omega$ are shown in
Figs.~8 and 9 for the A$\sim$130 and A$\sim$160 mass regions, respectively.

In Refs.~\cite{[Liu96],[Liu95]}, the spin assignments
for the A$\sim$130 and A$\sim$160 nuclei were investigated by energy
systematics. Our assignments are in general consistent with their
results. However, Liu {\it et al.} give three different 
sets of assignments for the Cs isotopes 
depending on the choice of the
reference nucleus\cite{[Liu96]}.
One of the sets coincides with our result. 
Systematical changes in the assigned spin values (e.g. by 2$\hbar$)
do not change the (energy) systematics \cite{[Liu96]}.
We have chosen $^{128}$La\cite{[Hay95]} and the 
lighter isotope of $^{120}$Cs~\cite{[Ced92]}
guided by theoretical results, see Fig.~8.
This assignment is also consistent with the measurements
for $^{130}$Cs\cite{[Sal91]}.
The nice agreement between
the calculation and experiment (in particular for $^{128}$La) gives
some confidence for the present method, see Fig.~8.
Especially the low spin-part agrees rather well with the experimental
data. The increase in angular momentum with frequency
for the favored signature is in general too strong when
compared to experiment.

Fig.~10 depicts the equilibrium deformation parameters 
$\beta_2$ and $\gamma$  
obtained from the TRS calculations. 
For the $A\sim 130$ nuclei, in general, the $\beta_2$ and
$\gamma$ deformations decrease with increasing neutron number.
However, the nuclear shapes of some $A\sim 130$ nuclei
are usually rather soft, which may lead to some inaccuracy
in determining the equilibrium deformations as well as 
it may affect other observables, in particular $I_x$ values.
The calculated deformations of the two signatures are rather close, but
one should be aware that this may not always be the case,
see e.g $^{122}$Cs (Fig.~10).

For the $A\sim 160$ isotopes, the 
the $\beta_2$ ($\gamma$) deformation value is
increasing (decreasing) with
neutron number, $N$. This is consistent
with the increase of the moments of inertia with $N$.
The $\gamma$ deformations of the $A\sim 160$
nuclei are rather small. With increasing $N$,
the $\gamma$ deformations are predicted to change from small positive to 
small negative values.
For most cases, the calculations yield a decrease of the deformation 
with increasing rotational frequency.

The quality of the present calculations is nicely demonstrated
in Fig.~11 and Fig.~12, where we 
compare experimental and calculated Routhians, $e^\prime$.
Especially for the Cs- and La-siotopes,
the agreement is in general within 50~keV. For the case of the heavier
Rare Earth nuclei, 
at relatively high
frequency $\hbar\omega\ge 0.25$\,MeV, the calculated Routhians
do not agree well with the experimental data. 
The deviations between experiment and theory can be linked to the
fact that the unblocked neutron $i_{13/2}$ crossings occur too
early in the calculations for some cases. 

We also compare the difference of the Routhians 
$\Delta e^\prime =e^\prime_f- e^\prime_u$ obtained in the TRS-calculations
and experiment, Fig.~13 and Fig.~14.
\footnote{Note that here 
the sign of $\Delta e^\prime$
is opposite to the one chosen in previous figures}
In order to calculate the
experimental values of $\Delta e^\prime (\hbar\omega)$ we performed
a linear interpolation of the unfavoured Routhians $e^\prime_u$ 
at the frequency values of the favoured signature.
Again, the agreement for the A=120-130 region is quite good.
On the other hand, for the Rare Earth nuclei, the inversion is calculated
to continue to somewhat higher frequencies and also
to be more pronounced than in experiment. 
However, given
the rather modest values of the inversion, in general
$<30$~keV, one can be quite content with the results. 
Above the frequency
of $\hbar\omega =0.5$~MeV (not seen in Fig.~14), 
the inversion has disappeared for
all cases considered here. 
The TRS-calculations clearly show that the mean-field
model can account for the signature inversion phenomenon,
once shape-polarization and pairing effects are treated
self-consistently. The anomalous signature 
splitting is obtained already at almost
axial shapes, due to the contribution of the $(\lambda\mu)=(22)$ 
$QQ$-pairing interaction to the single-particle potential.
The overall agreement between theory and experiment appears
satisfactory.

\section{Summary}

We have demonstrated that shape and pairing self-consistent 
mean-field calculations can account 
for the rotational band structure of the unique parity high-$j$
orbits in the
odd-odd A$\sim$130 and A$\sim$160 nuclei.
In particular, the signature inversion
phenomenon is reproduced. 
The agreement between theory and the data is satisfactory 
provided that some of the empirical spin assignments are revised.
New spins are suggested based 
on a systematic analysis of the \jone~moments of inertia which rather firmly
establishes relative spins along each isotopic chain considered here.  
The absolute spin values are finally determined
partly guided (see Sec.~IV) by theoretical values.

The experimental 
signature-splittings are quite well reproduced although the TRS 
calculations yield rather modest values of the triaxiality parameter
$\gamma$ ($\gamma<5^\circ$) for most cases. This can be understood
in terms of the additional 
enhancement of the anomalous signature splitting caused by the mean-field
contribution of the  $(\lambda\mu)=(22)$ component of the 
quadrupole pairing interaction, see Sec.~III.

Some deficiencies of the model are clearly visible particularly
at high rotational frequency and for $\gamma$-soft nuclei.
Other effects, such as the coupling between rotation and $\gamma$-vibration, 
may need to be taken into account.
Also more elaborate forces, 
in particular accounting for the valence neutron-proton interaction
\cite{[Ham86b],[Sem91]}, might further improve the agreement to the data.
For a full understanding of this intriguing phenomenon, the 
calculations of electromagnetic transitions rates are important.
This, however, is beyond the scope of our work.

\bigskip

  This work was supported by the Swedish Institute (SI), Swedish Natural
Science Research Council (NFR), the G\"oran Gustafsson Foundation
and the Polish Committee for Scientific
Research (KBN) under Contract No. 2~P03B~040~14.

\bigskip


\newpage

\begin{table}
\label{Table.I}
\caption[]{Assigned spins and parities for the lowest observed levels
of the $\pi h_{11/2}\bigotimes\nu h_{11/2}$ bands in the Cs and La isotopes
and the $\pi h_{11/2}\bigotimes\nu i_{13/2}$ bands in the A$\sim$160 nuclei.
The references for the previous assignments are shown in brackets.} 
\begin{tabular}{lllllllll}
Nuclei & $^{120}$Cs & $^{122}$Cs & $^{124}$Cs & $^{126}$Cs
& $^{128}$Cs & $^{130}$Cs & & \\
Previous & $8^+$\cite{[Ced92]} & $6^-$\cite{[Xu90]}$^a$
& $7^+$\cite{[Kom93]} & $5^+$\cite{[Kom93]}
& $9^+$\cite{[Pau89]} & $9^+$\cite{[Sal91]} & & \\
Present & $8^+$ & $9^+$ & $9^+$ & $9^+$ & $10^+$ & $9^+$ & & \\
\hline
Nuclei & $^{124}$La & $^{126}$La & $^{128}$La & $^{130}$La & $^{132}$La
& $^{134}$La & &\\
Previous & $7^+$\cite{[Liu96]} & $4^+$\cite{[Nya89]}
& $5^+$\cite{[Hay95]} & $8^+$\cite{[Pau87]} & $8^+$\cite{[Oli89]}
& $8^+$\cite{[Oli92]} & & \\
Present & $7^+$ & $7^+$ & $5^+$ & $9^+$ & $9^+$ & $9^+$ & & \\
\hline
Nuclei & $^{154}$Tb & $^{156}$Tb
& $^{156}$Ho& $^{158}$Ho & $^{160}$Ho& $^{158}$Tm
& $^{160}$Tm& $^{162}$Tm\\
Previous & $9^-$\cite{[Ben82]} & $6^-$\cite{[Ben82]}
& $9$\cite{[Hel92]} & $6^-$\cite{[Hel96]} & $6^-$\cite{[Sal90]}
& $9^-$\cite{[Dri86]} & $9^-$\cite{[Dri86]}  & $7^-$\cite{[Dri86]}  \\
Present  & $9^-$ & $8^-$ & $9^-$ & $9^-$ & $6^-$ & $9^-$ & $9^-$ & $7^-$ 
\end{tabular}
$^a$This band was assigned in Ref.~\cite{[Xu90]} as 
 $\pi h_{11/2}\bigotimes\nu g_{7/2}$ and later
 reassigned as the $\pi h_{11/2}\bigotimes\nu h_{11/2}$ configuration in
 Refs.~\cite{[Kom93],[Liu96]}. The present calculation supports the
 $\pi h_{11/2}\bigotimes\nu h_{11/2}$ configuration assignment.
\end{table}

\begin{figure}
\epsfig{file=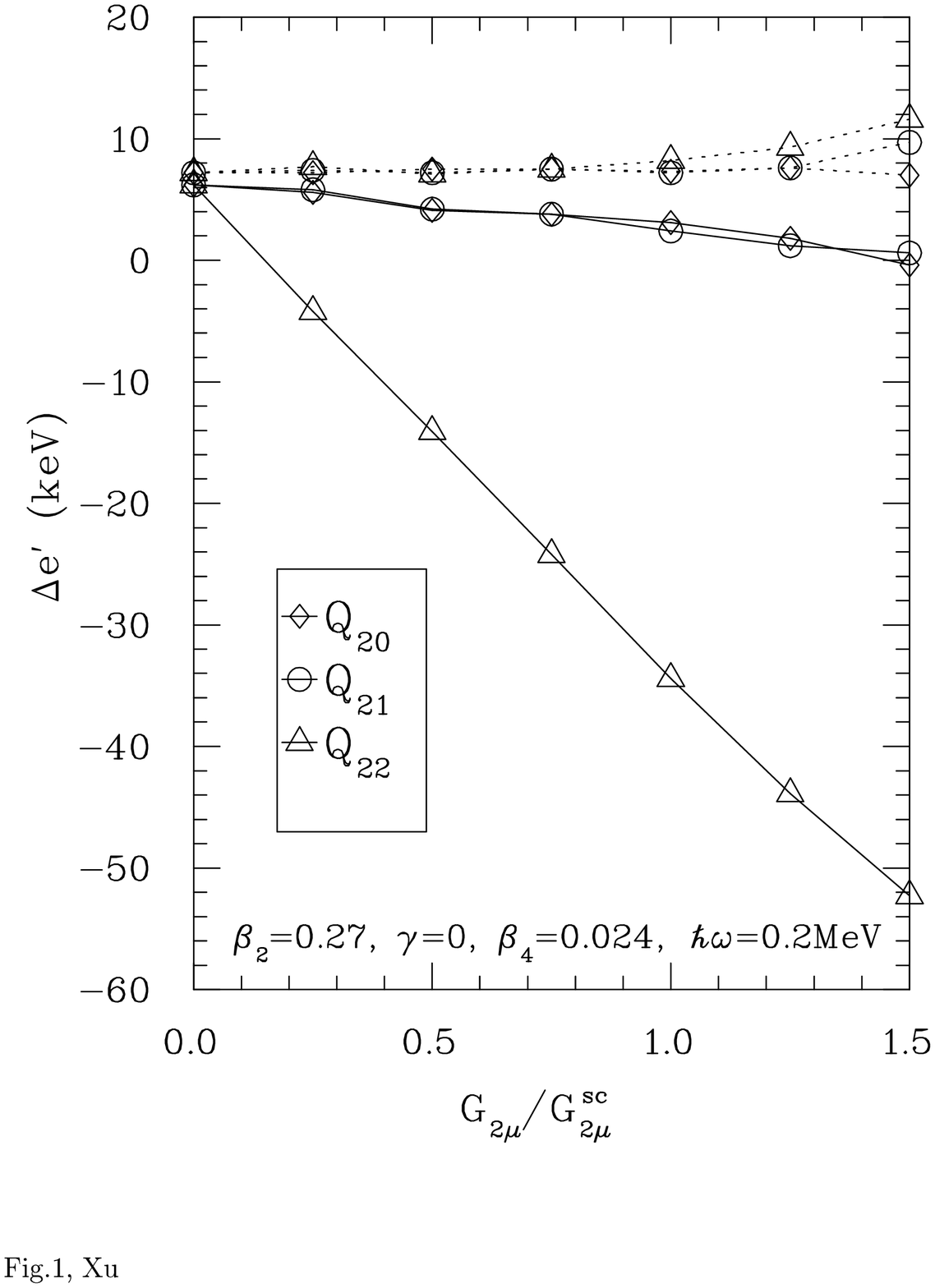,width=0.75\textwidth}
\label{Fig.1}
\caption[]{The contribution of each $Q_{2\mu}$ 
component to the signature splitting 
in $^{120}$Cs
as a function of the $QQ$-pairing force strength 
$G_{2\mu}/G_{2\mu}^{\rm sc}$, $(\mu=0,1,2)$.
The $G_{2\mu}^{\rm sc}$ is the strength of 
$QQ$-pairing determined according to the prescription of
Ref.~\cite{[Sak90]}. Calculations have been performed at 
$\hbar\omega$=0.20\,MeV. The $\beta_2$ and $\beta_4$ parameters 
correspond to the equilibrium deformation obtained in the TRS 
calculations but the $\gamma$ deformation was set
to zero in order to address the effects of the  $QQ$-pairing force alone.
The solid/dotted lines mark calculations with/without
the $\Gamma^{(2\mu )}$ potential.
The negative value of $\Delta e^\prime$ implies
the occurrence of signature inversion.}
\end{figure}

\begin{figure}
\epsfig{file=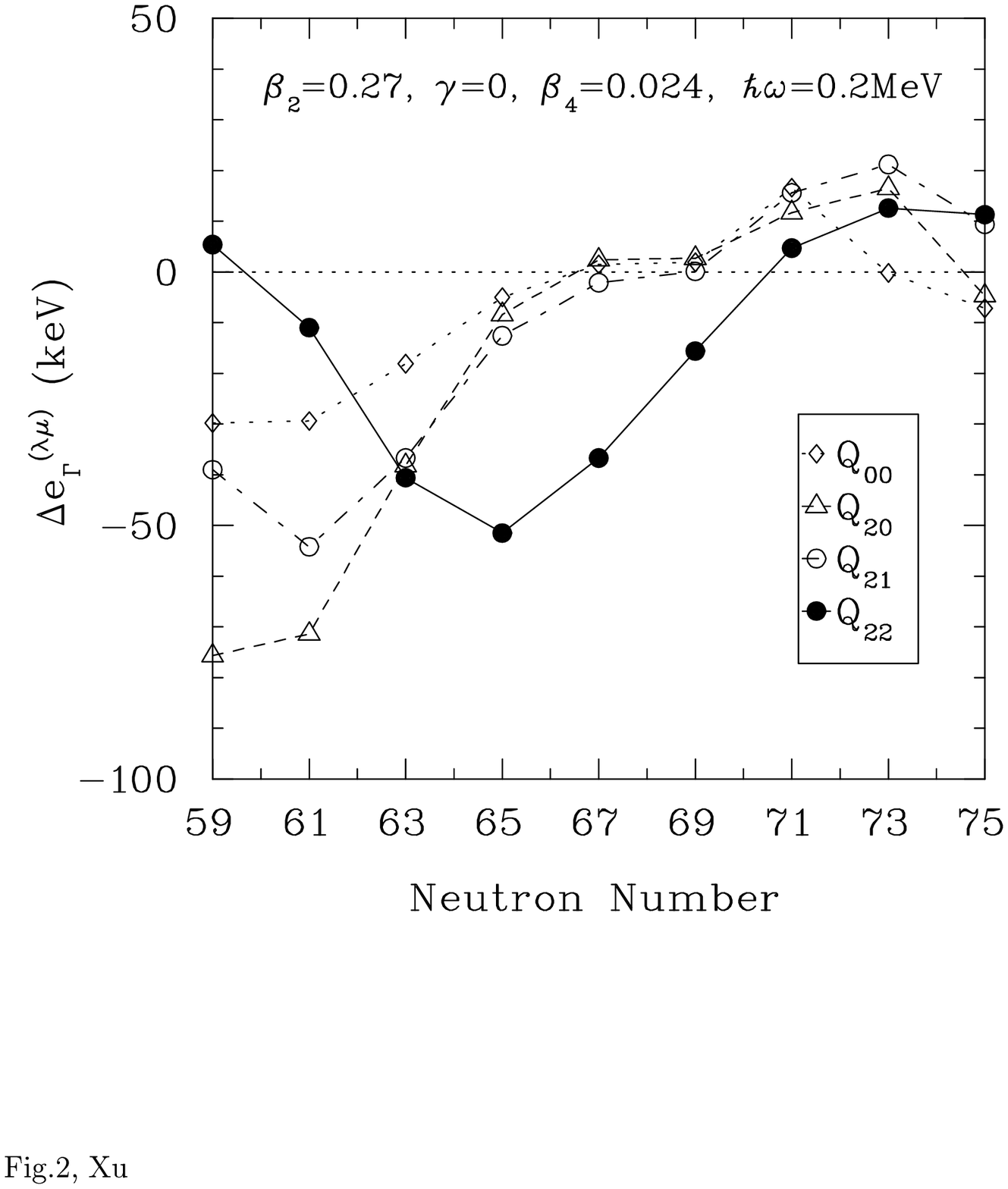,width=0.75\textwidth}
\vfill
\label{Fig.2}
\caption{The contribution of each component of the $\Gamma^{(\lambda\mu)}$ 
potential (see text for details) to the signature splitting in A$\sim$130 
nuclei as a function of neutron number. 
}
\end{figure}

\begin{figure}
\epsfig{file=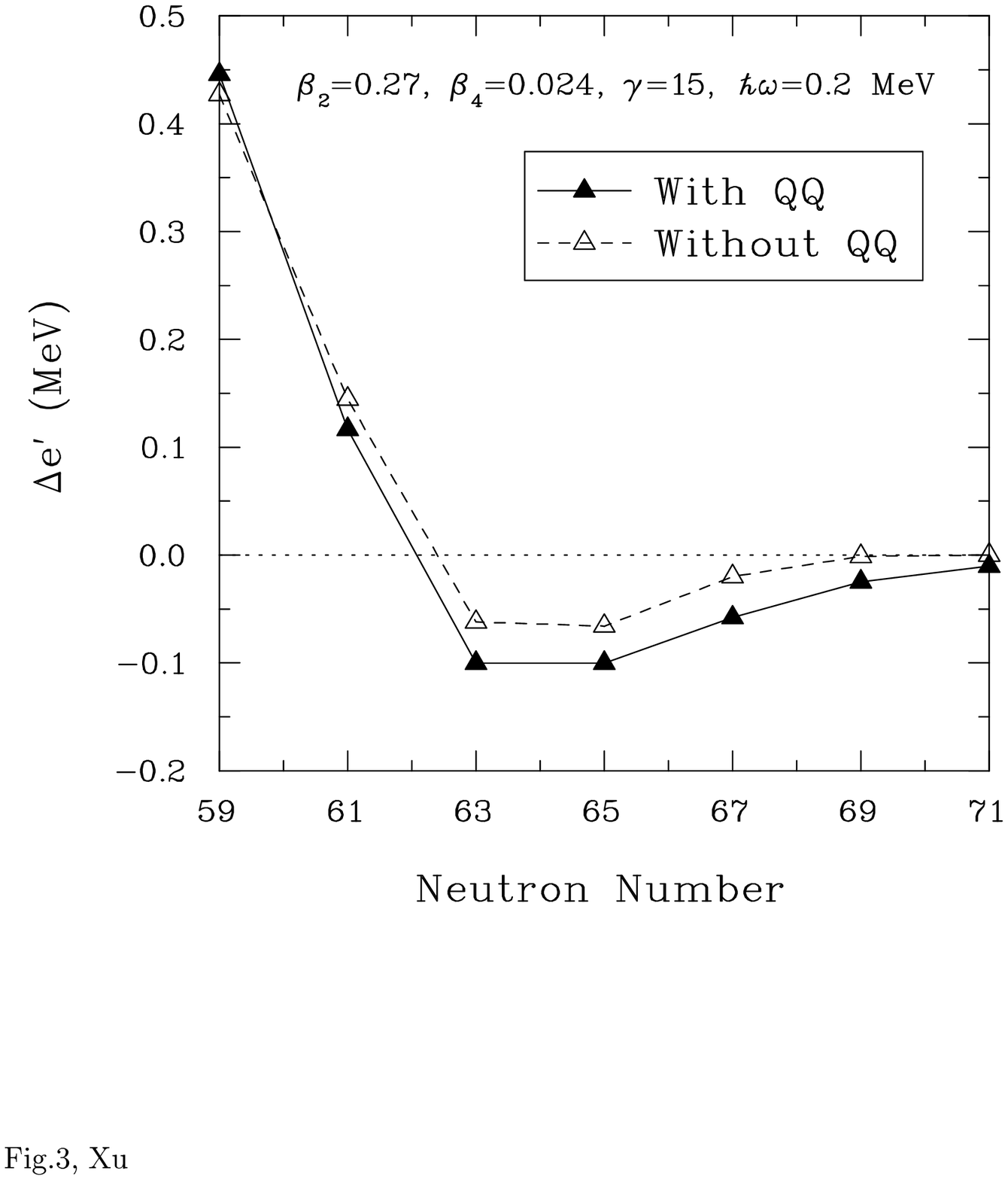,width=0.75\textwidth}
\label{Fig.3}
\vfill
\caption{Signature splitting between the unfavored and favored levels
of the $h_{11/2}$ quasi-neutron as  a function of neutron shell filling. 
Rotational
frequency and deformation (triaxial) were kept constant. 
Calculations with (without) the $QQ$-pairing interaction
are marked by solid (dashed) line and filled (open) triangles,
respectively.}
\end{figure}

\begin{figure}
\epsfig{file=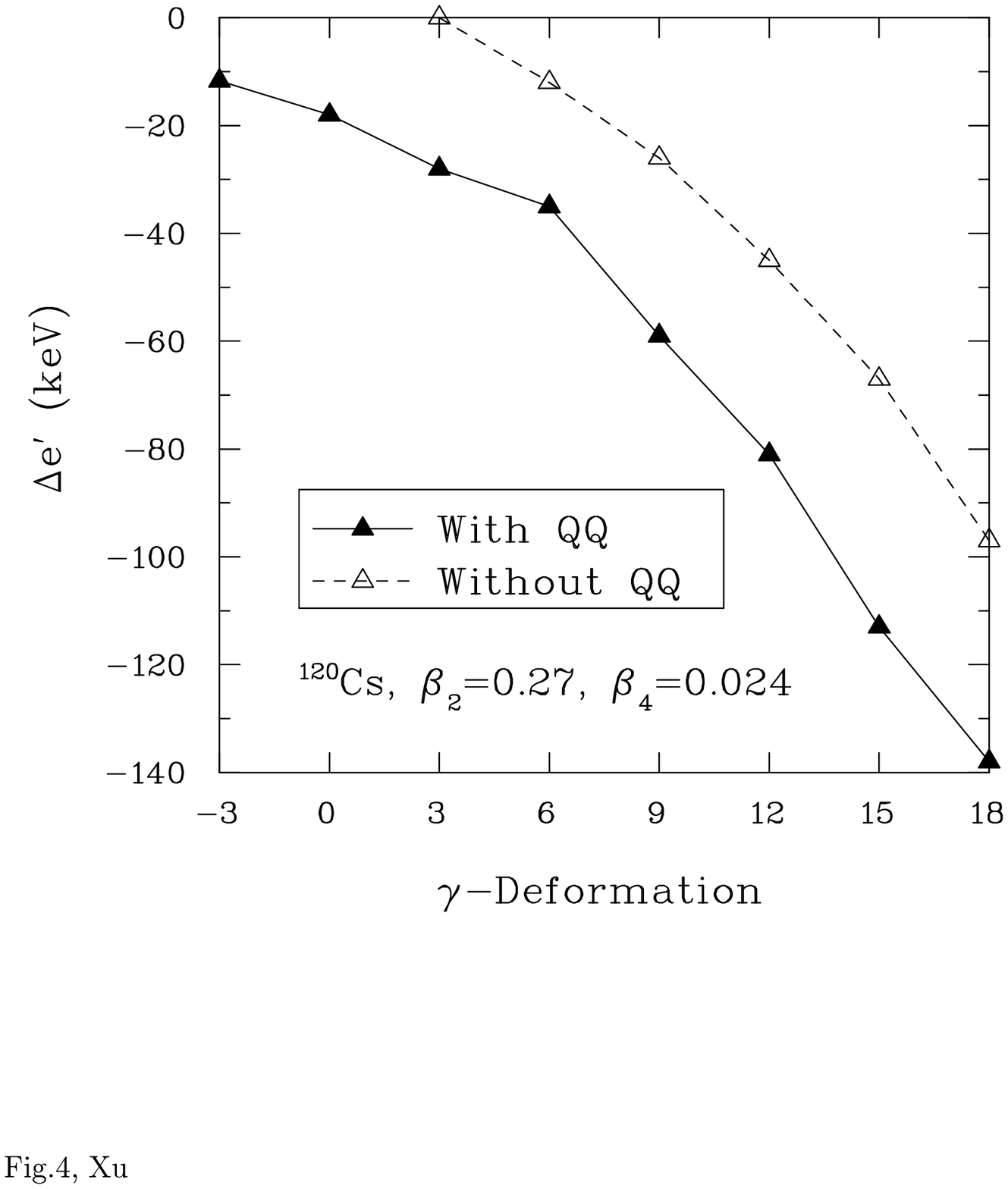,width=0.75\textwidth}
\label{Fig.4}
\vfill
\caption{The variation of the signature splitting versus
$\gamma$ deformation for $^{120}$Cs.}
\end{figure}

\begin{figure} 
\epsfig{file=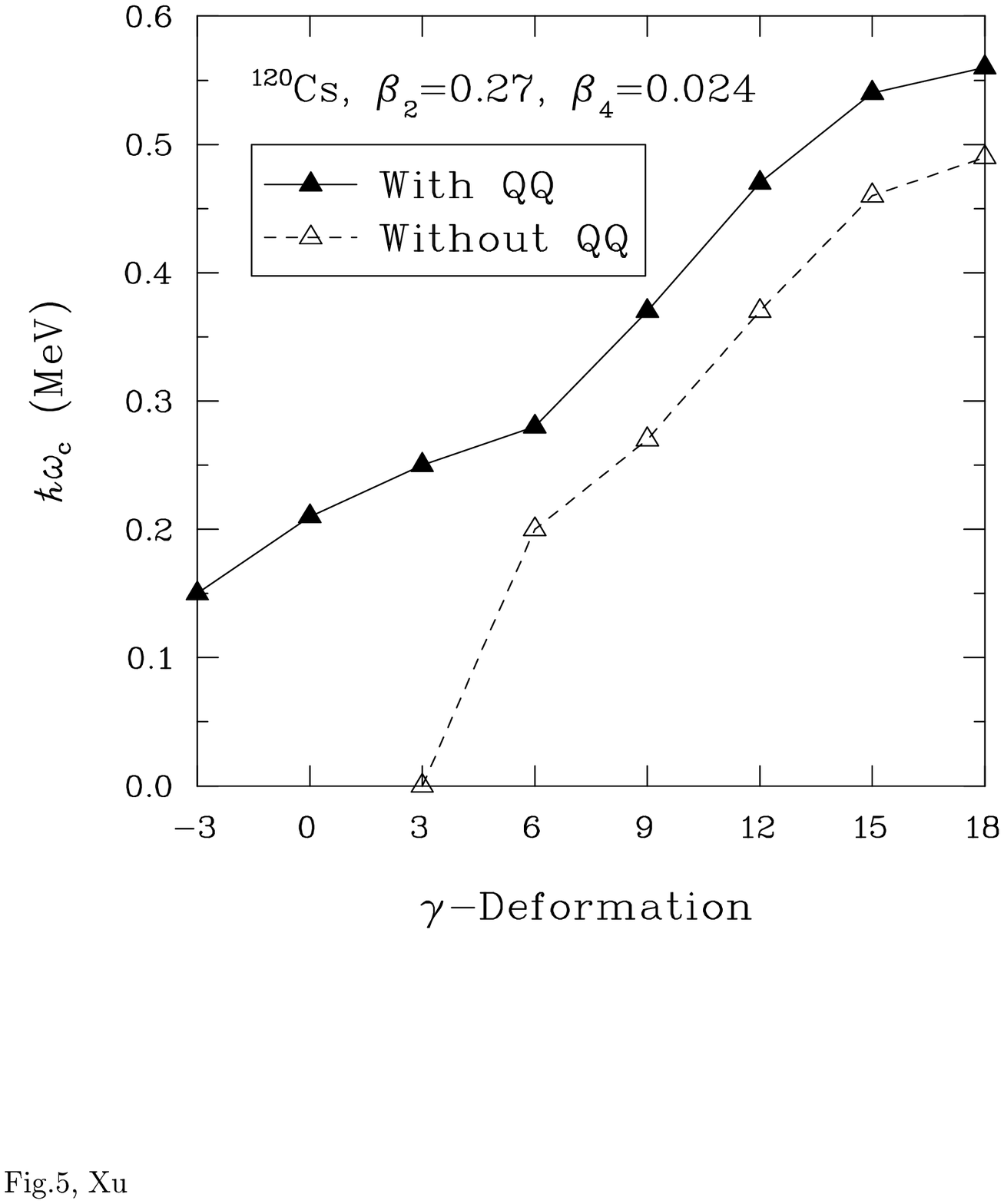,width=0.75\textwidth} \label{Fig.5}
\vfill
\caption{The critical frequency, $\hbar\omega_c$, at which the normal    
signature splitting is restored as a function of triaxiality $\gamma$.
} \end{figure}

\begin{figure}
\epsfig{file=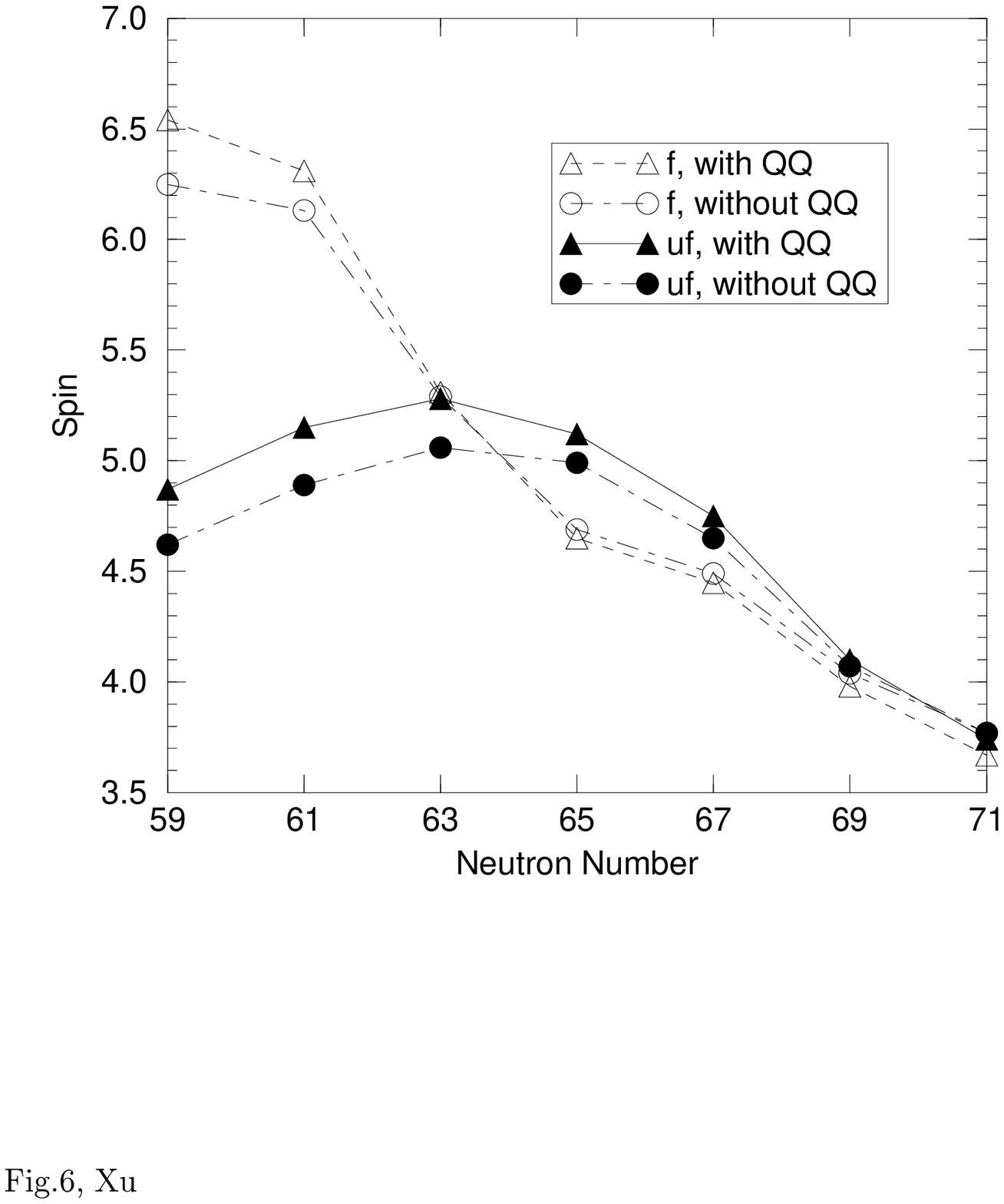,width=0.75\textwidth}
\label{Fig.6}
\vfill
\caption{The neutron contribution to the total angular momentum of
the favored (f) and unfavored (uf) signature bands as a function of
neutron number at $\hbar\omega=0.2$ MeV.
The deformation is set to the same values as in Fig.~3.
}
\end{figure}

\begin{figure}
\epsfig{file=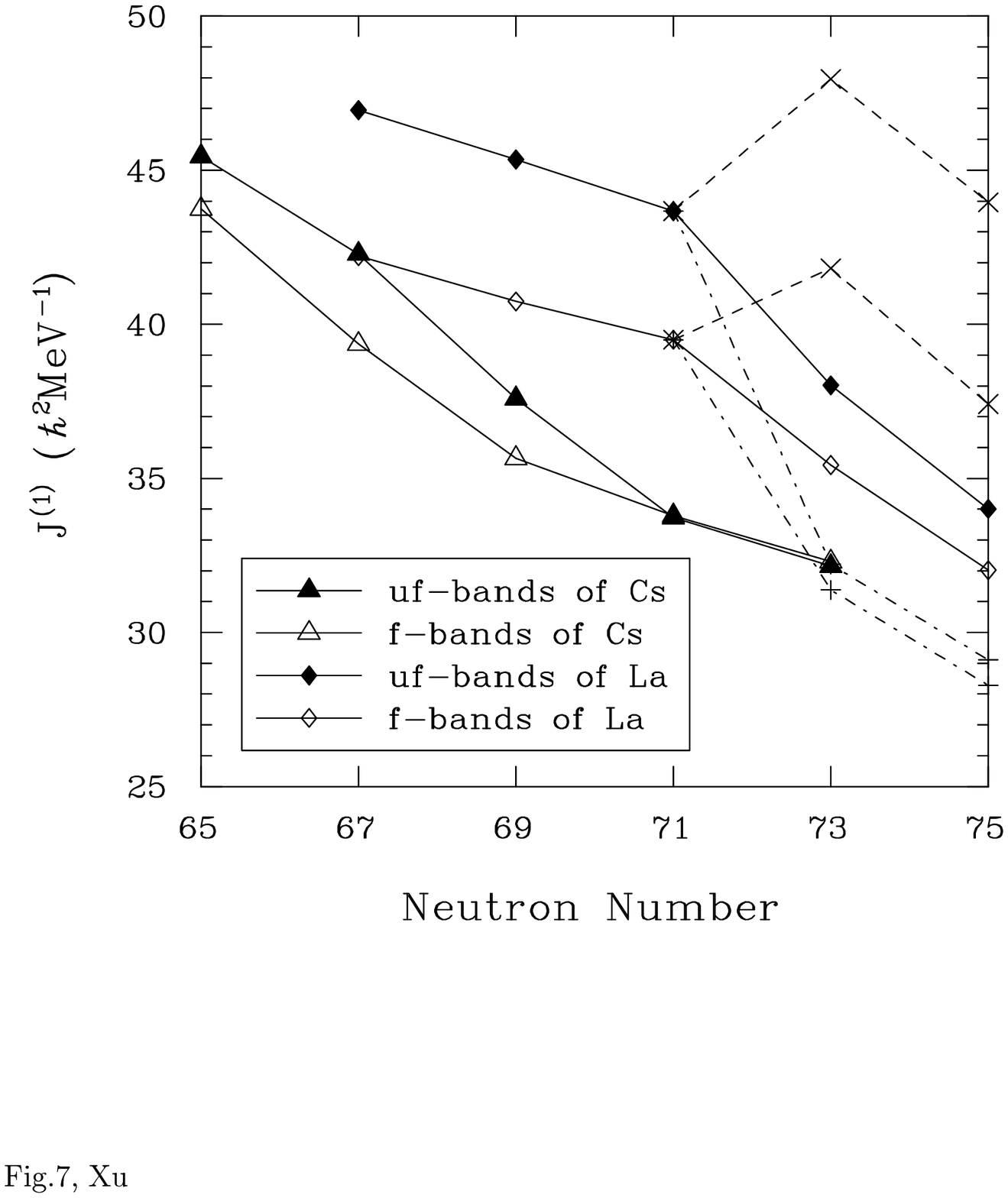,width=0.75\textwidth}
\label{Fig.7}
\vfill
\caption{The experimental moments of inertia
for the Cs (filled symbols) and La (open symbols) isotopes
corresponding to our spin assignments following the discussion in Sec.~IV.
For the favored (unfavored) bands, 
the $\jone$~values at $I=11(10)\hbar$ are plotted, respectively.
The dashed (dot-dashed) curves indicate
variations in $\jone$~of $^{130, 132}$La caused by changes in 
the spin values by $+1(-1)\hbar$, 
respectively.}
\end{figure}

\begin{figure}
\epsfig{file=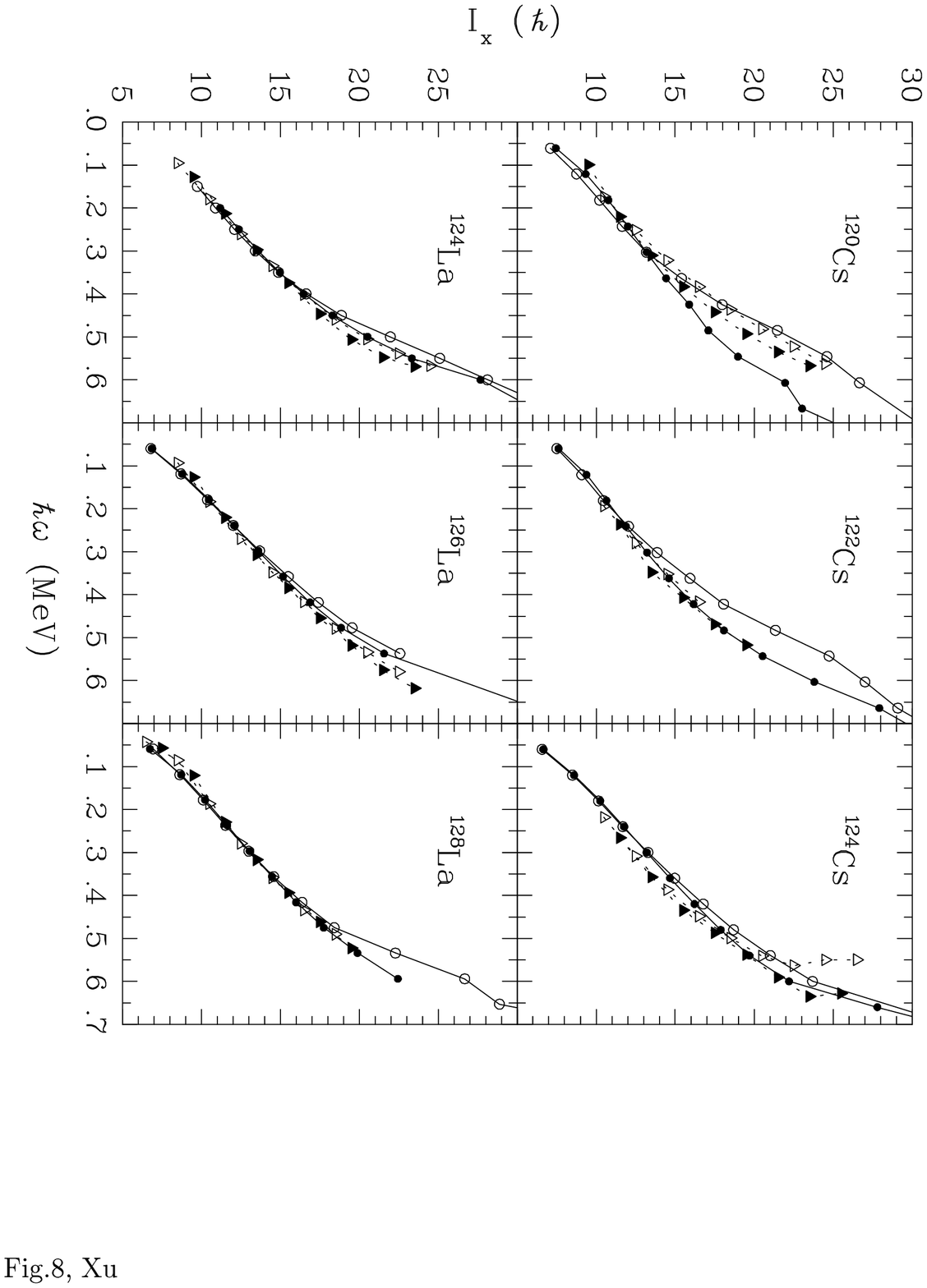,width=0.75\textwidth}
\label{Fig.8}
\vfill
\caption{Experimental and calculated total angular momenta ($I_x$) for
the $\pi h_{11/2}\bigotimes\nu h_{11/2}$ bands in odd-odd $^{120-124}$Cs and
$^{124-128}$La nuclei. The triangles (empty and filled) correspond to the
experimental data (favored and unfavored signatures, respectively).
The circles (empty and filled) mark
the TRS results (favored and unfavored signatures, respectively).
Our new spin assignments (Sec.~IV) are used. The references to 
the original data are found in Table~I.}
\end{figure}

\begin{figure}
\epsfig{file=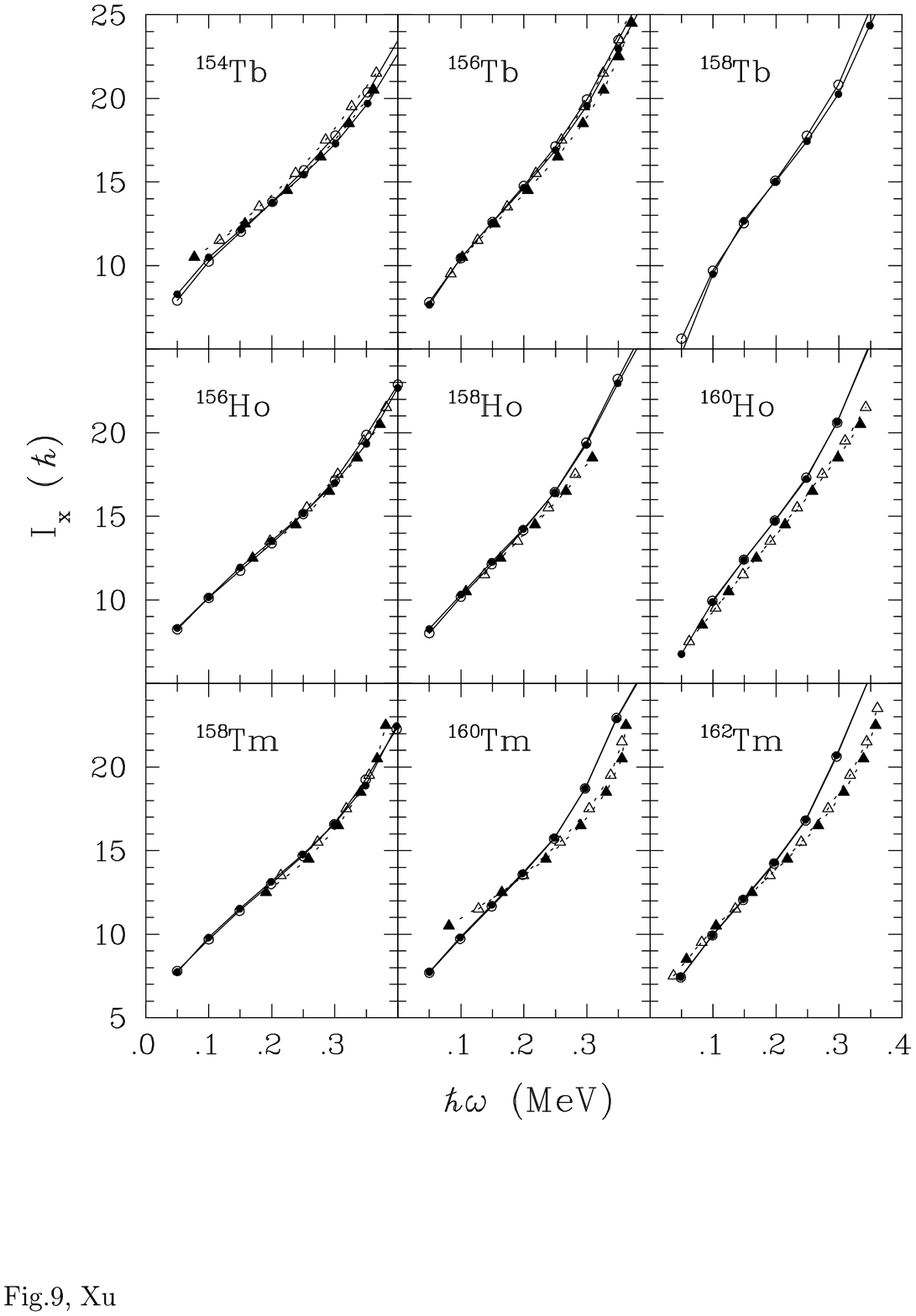,width=0.75\textwidth}
\label{Fig.9}
\vfill
\caption{Similar to Fig.~8, but for  the
$\pi h_{11/2}\bigotimes\nu i_{13/2}$ bands in the A$\sim$160 nuclei.}
\end{figure}

\begin{figure}
\epsfig{file=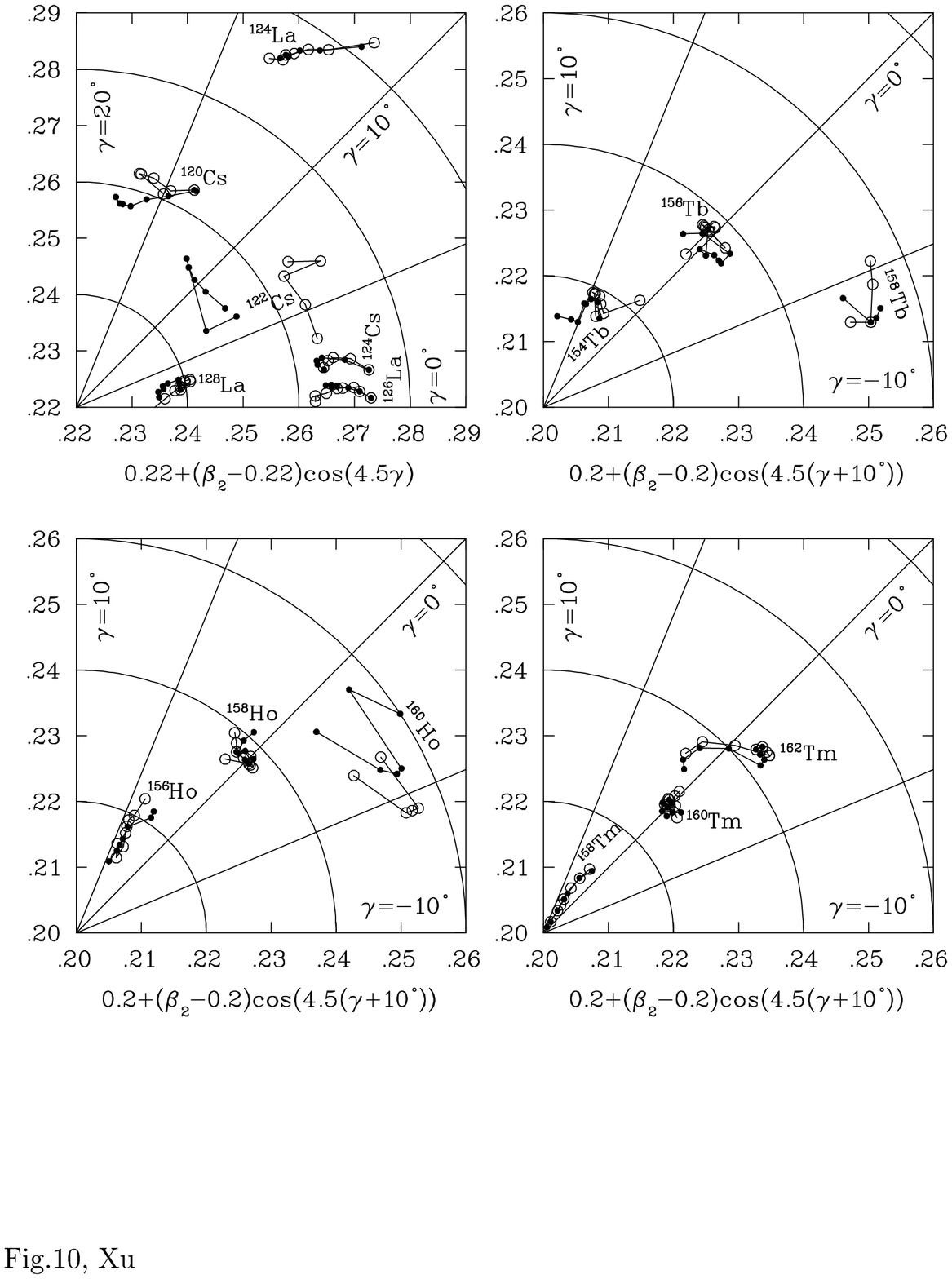,width=0.75\textwidth}
\label{Fig.10}
\vfill
\caption{The calculated quadrupole equilibrium deformations $\beta_2$ and 
$\gamma$ for the $\pi h_{11/2}\bigotimes\nu h_{11/2}$
and $\pi h_{11/2}\bigotimes\nu i_{13/2}$ rotational bands.
The empty (filled) circles are for favored (unfavored) bands, respectively.}
\end{figure}

\begin{figure}
\epsfig{file=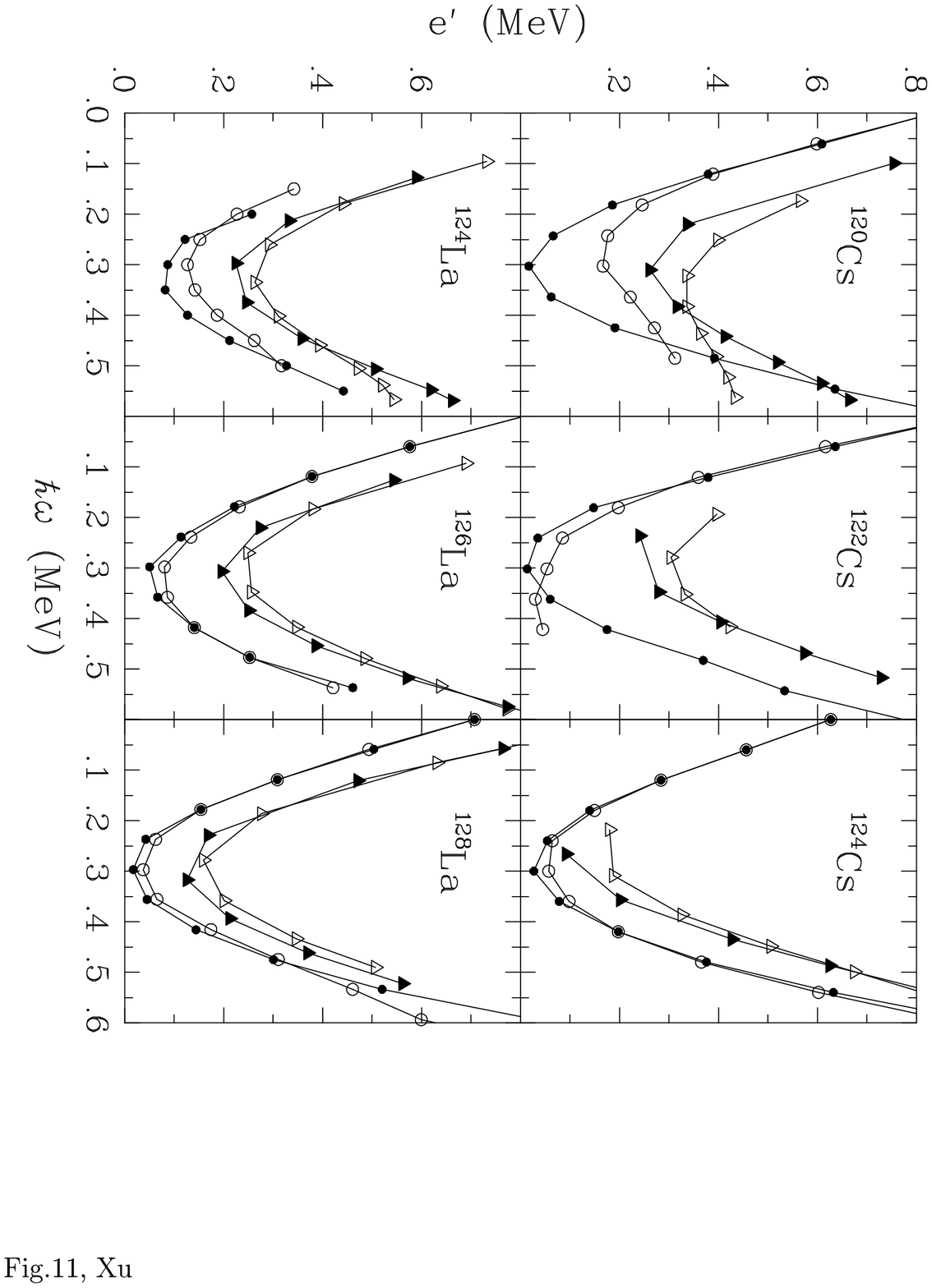,width=0.75\textwidth}
\label{Fig.11}
\vfill
\caption{Experimental and calculated Routhians for the 
$\pi h_{11/2}\bigotimes\nu h_{11/2}$ bands in Cs
and La isotopes as a function of rotational frequency. 
A common reference Routhian of
$J_0=44$\,$\hbar^2$MeV$^{-1}$ 
has been subtracted.
The triangles (empty and filled) mark
the experimental data (favored and unfavored signatures, respectively),
while (empty and filled) circles denote
the TRS results (favored and unfavored signatures, respectively).
The assignments of Sec.~IV have been used. 
Note, that the normalization of the Routhians is arbitrary.
}
\end{figure}

\begin{figure}
\epsfig{file=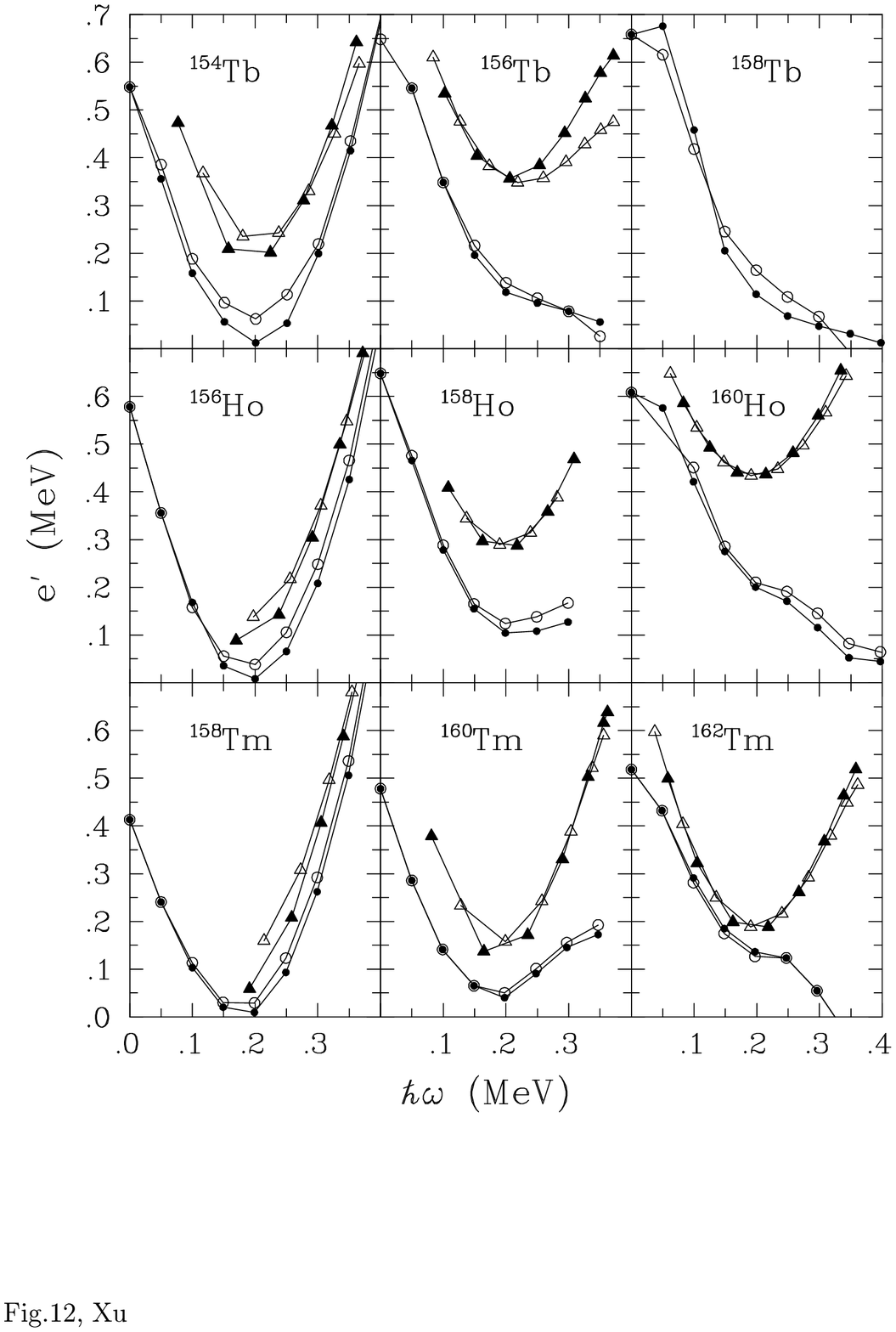,width=0.75\textwidth}
\label{Fig.12}
\vfill
\caption{Similar to Fig.~11, but for the 
$\pi h_{11/2}\bigotimes\nu i_{13/2}$ bands in A$\sim$160 nuclei.
The inertia parameter of the subtracted reference Routhian is
$J_0=70$\,$\hbar^2$MeV$^{-1}$.}
\end{figure}

\begin{figure}
\epsfig{file=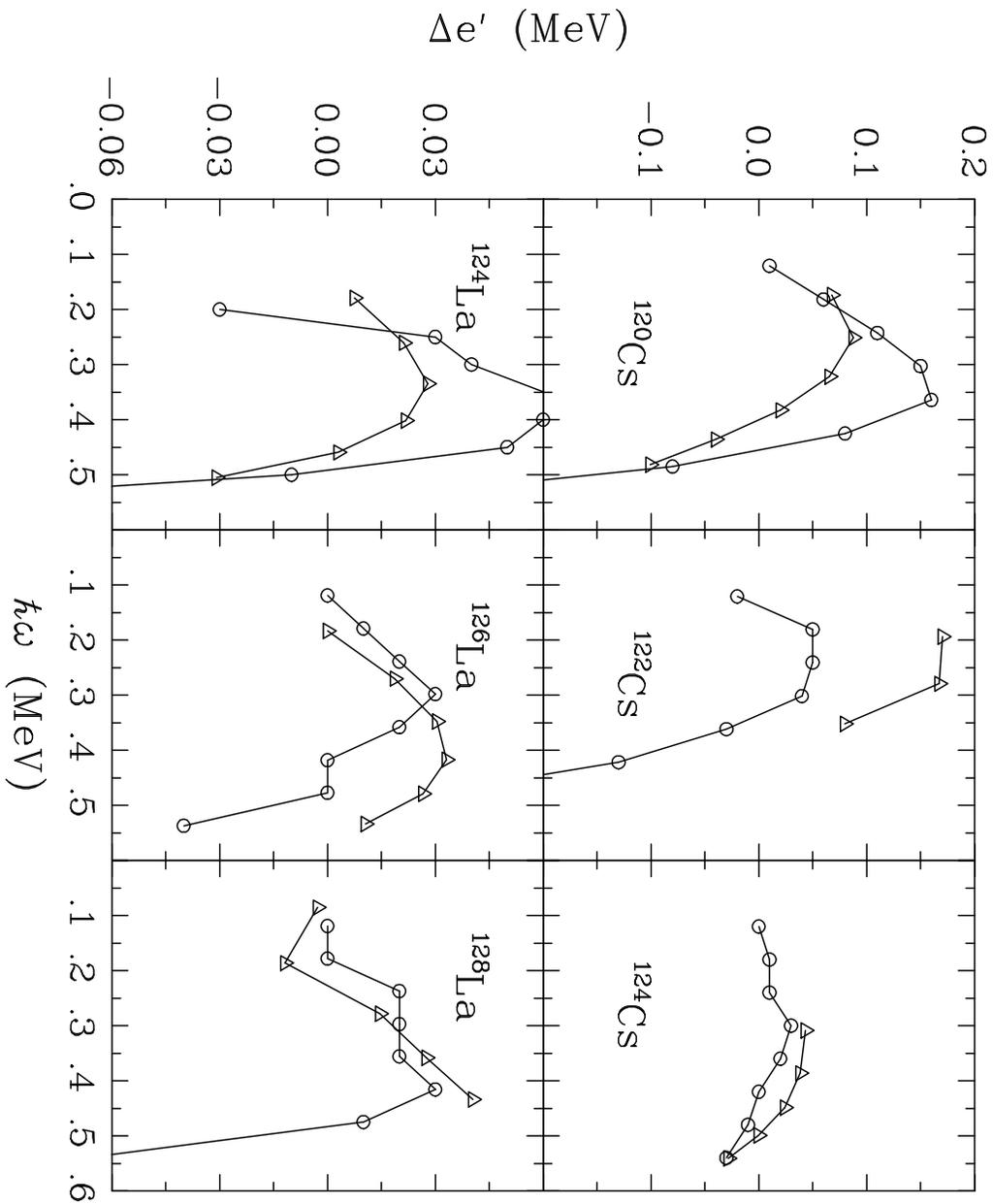,width=0.75\textwidth}
\label{Fig.13}
\vfill
\caption{The difference of calculated and experimentally
deduced Routhians from Fig.~11, $\Delta e^\prime=e^\prime _f - e^\prime _u$,
where $f~(u)$ stands for (un-)favoured signature. Positive values of
$\Delta e^\prime$ correspond to signature inversion. 
The triangles (circles) characterize experimental (calculated) values. 
Note that the scales for the Cs- and La-isotopes are different.}
\end{figure}

\begin{figure}
\epsfig{file=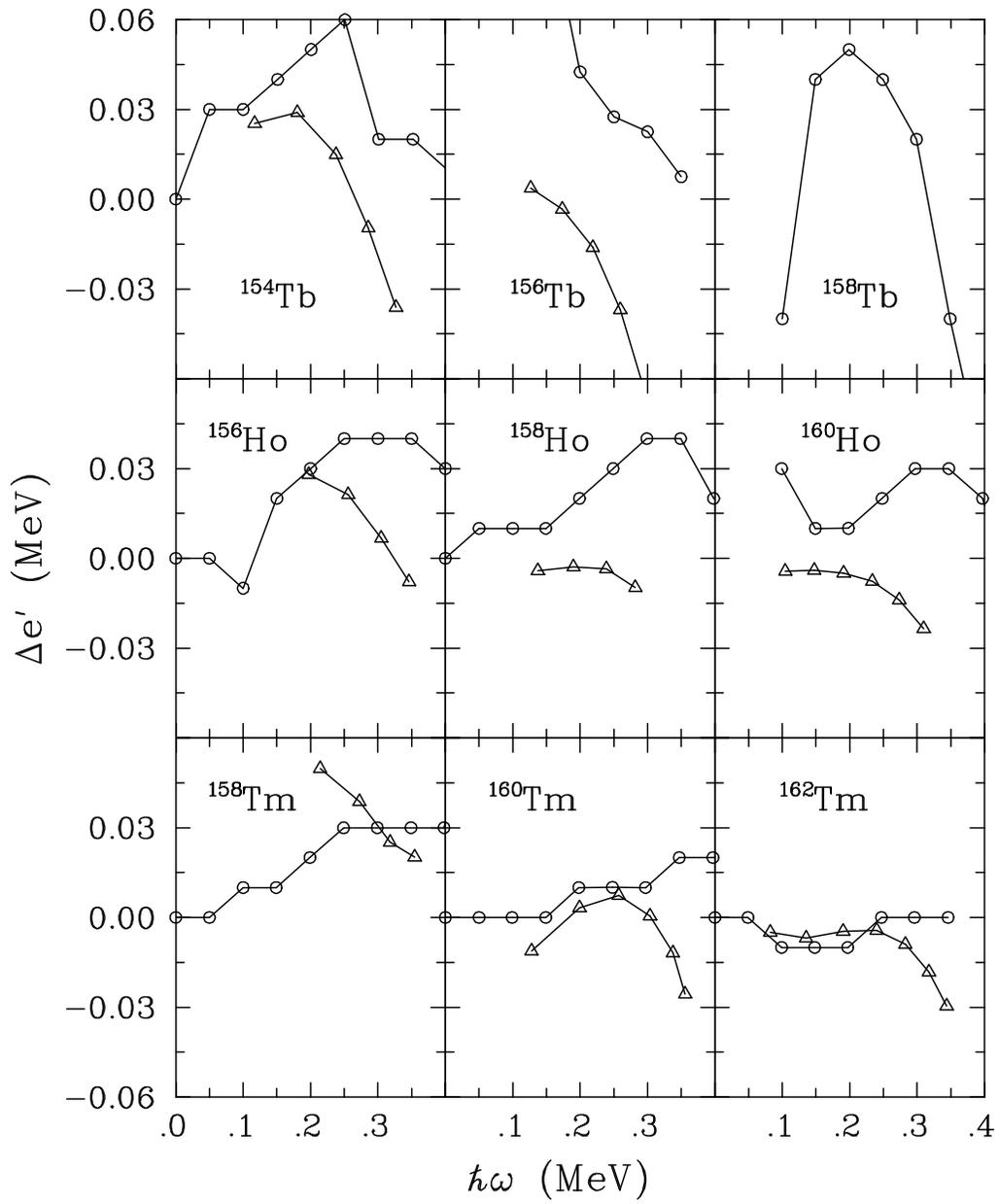,width=0.75\textwidth}
\label{Fig.14}
\vfill
\caption{Similar to Fig.~13, but for the 
$\pi h_{11/2}\bigotimes\nu i_{13/2}$ bands in A$\sim$160 nuclei.
}
\end{figure}
\end{document}